\documentclass{Interspeech2024}

\interspeechcameraready 
\usepackage{times}
\usepackage{latexsym}
\usepackage{graphicx}
\usepackage{float}
\usepackage{amssymb}
\usepackage{mathtools}
\usepackage{amsthm}
\usepackage{amsfonts}
\usepackage{subfig}
\usepackage{caption}
\usepackage{subcaption}
\usepackage{amsmath}
\usepackage{algorithm2e}
\usepackage{multirow} 
\usepackage{balance}
\usepackage{cite}

\usepackage[T1]{fontenc}

\title{VECL-TTS: Voice identity and Emotional style controllable Cross-Lingual Text-to-Speech}

\name[affiliation={1}]{Ashishkumar}{Gudmalwar}
\name[affiliation={1}]{Nirmesh}{Shah}
\name[affiliation={1}]{Sai}{Akarsh}
\name[affiliation={1}]{Pankaj}{Wasnik}
\name[affiliation={2}]{Rajiv Ratn}{Shah}

\address{
  $^1$Sony Research India Pvt. Ltd., Bangalore, India\\
  $^2$Indraprastha Institute of Information Technology (IIIT), Delhi, India}
  
\email{\{ashish.gudmalwar1,nirmesh.shah,pankaj.wasnik\}@sony.com, rajivratn@iiitd.ac.in}

\begin{document} 
\maketitle
\begin{abstract}
Despite the significant advancements in Text-to-Speech (TTS) systems, their full utilization in automatic dubbing remains limited. This task necessitates the extraction of voice identity and emotional style from a reference speech in a source language and subsequently transferring them to a target language using cross-lingual TTS techniques. While previous approaches have mainly concentrated on controlling voice identity within the cross-lingual TTS framework, there has been limited work on incorporating emotion and voice identity together. To this end, we introduce an end-to-end Voice Identity and Emotional Style Controllable Cross-Lingual (VECL) TTS system using multilingual speakers and an emotion embedding network. Moreover, we introduce content and style consistency losses to enhance the quality of synthesized speech further. The proposed system achieved an average relative improvement of 8.83\% compared to the state-of-the-art (SOTA) methods on a database comprising English and three Indian languages (Hindi, Telugu, and Marathi).

\end{abstract}
\section{Introduction}
With the advancement of sophisticated Text-to-Speech (TTS) systems, research has notably shifted towards implementing speech-generation technologies for automatic dubbing applications \cite{liu4564500m}. Particularly, Cross-lingual TTS plays a vital role in automatic dubbing by generating high-quality speech in a target language while upholding the distinctive voice identity and emotional nuances of the original speaker. The core idea involves extracting the vocal characteristics of a speaker in the source language, specifically English, and transferring them to a foreign language, such as an Indian language, or vice versa \cite{sun2016personalized,zhao2021towards,xin2021disentangled,xin2020cross,xie2016kl,himawan2020speaker}. This creates an impression that the English-speaking actor/actress is now conversing in an Indian language post-dubbing, which enhances the viewers' immersive experience.  

While the input text provided to the TTS may contain certain emotional cues, it often falls short of capturing the nuanced speaking style specific to a target speaker. Consequently, TTS systems typically generate speech based on the styles they have been trained on. Various methodologies have been proposed to address these limitations, aiming to extract voice identity style information from a reference speech signal \cite{wang2023valle, casanova2022yourtts, lancucki2021fastpitch}. However, many of these approaches assume both the reference speech signal and the target text for synthesis are in the same language. Furthermore, for emotion control, some techniques leverage one-hot vector embeddings or emotional features extracted from a reference speech signal in the same language \cite{wu2021cross,im2022emoq,shah2023nonparallel,singh23_interspeech}. More recently, attempts have been made to control voice identity using reference signals from different languages, a concept referred to as cross-lingual TTS \cite{casanova2022yourtts, zhang2019learning,li2024styletts}. Nevertheless, these attempts have not extensively delved into the control of emotion within a cross-lingual framework. Only one approach, named METTS \cite{zhu2023metts}, attempts to achieve multilingual Text-to-Speech (TTS) by incorporating cross-lingual emotion and cross-speaker transfer. METTS stands out due to its base architecture and the methodology employed for extracting and aligning emotion-embedding information from the reference signal. However, it's noteworthy that the emotion similarity of METTS is not sufficiently competitive when compared to state-of-the-art (SOTA) models \cite{zhu2023metts}.

In this paper, we propose a novel approach for the simultaneous control of voice identity and emotional style within a cohesive framework, aiming to develop an end-to-end system for Voice Identity and Emotional Style Controllable Cross-Lingual Text-to-Speech (TTS). Building upon the established VITS/YourTTS architecture \cite{casanova2022yourtts}, our design serves as an extension. To address a significant limitation in VITS/YourTTS, specifically the instability in predicting phoneme duration through the stochastic duration predictor, resulting in the production of unnatural speech, we introduce a conditioning mechanism for the stochastic duration network predictor. This involves utilizing both speaker and emotion embedding features, recognizing that duration is influenced by both styles.

Furthermore, we observe a substantial decline in the pronunciation quality of the generated speech after applying cross-lingual voice identity and emotion style, primarily due to unnatural duration. Therefore, we propose the application of content loss by integrating wav2vec2-based self-supervised speech representations \cite{baevski2020wav2vec} during training to mitigate pronunciation errors. Our findings indicate that the proposed model achieves better subjective and objective scores, particularly in emotion similarity, compared to state-of-the-art algorithms and various ablation studies across English and three Indian languages: Hindi, Telugu, and Marathi (these languages are spoken by more than 600 million people \cite{patil2013syllable}).

\noindent The key contributions can be summarized as follows:
\begin{itemize} 
    \item The proposed approach employs two separate multilingual speaker and emotion embedding extractor networks to achieve cross-lingual speaker and emotion transfer in a unified framework.
    \item Both speaker and emotion embeddings were employed to produce stable phoneme duration via a stochastic duration predictor network.
    \item We introduce emotion consistency loss to improve emotion controllability further.
    \item Wav2vec2 based self-supervised speech representations were utilized to reduce pronunciation errors via proposed content loss after cross-lingual transfer.
    \item This paper focuses on cross-lingual transfer among English and three Indian languages (Hindi, Telugu, and Marathi), which has received limited attention until now.
\end{itemize}
\section{Methodology}
This section discusses the proposed VECL-TTS architecture along with the motivation for different proposed components.
\begin{figure*}
    \centering
    \includegraphics[width=0.75\linewidth]{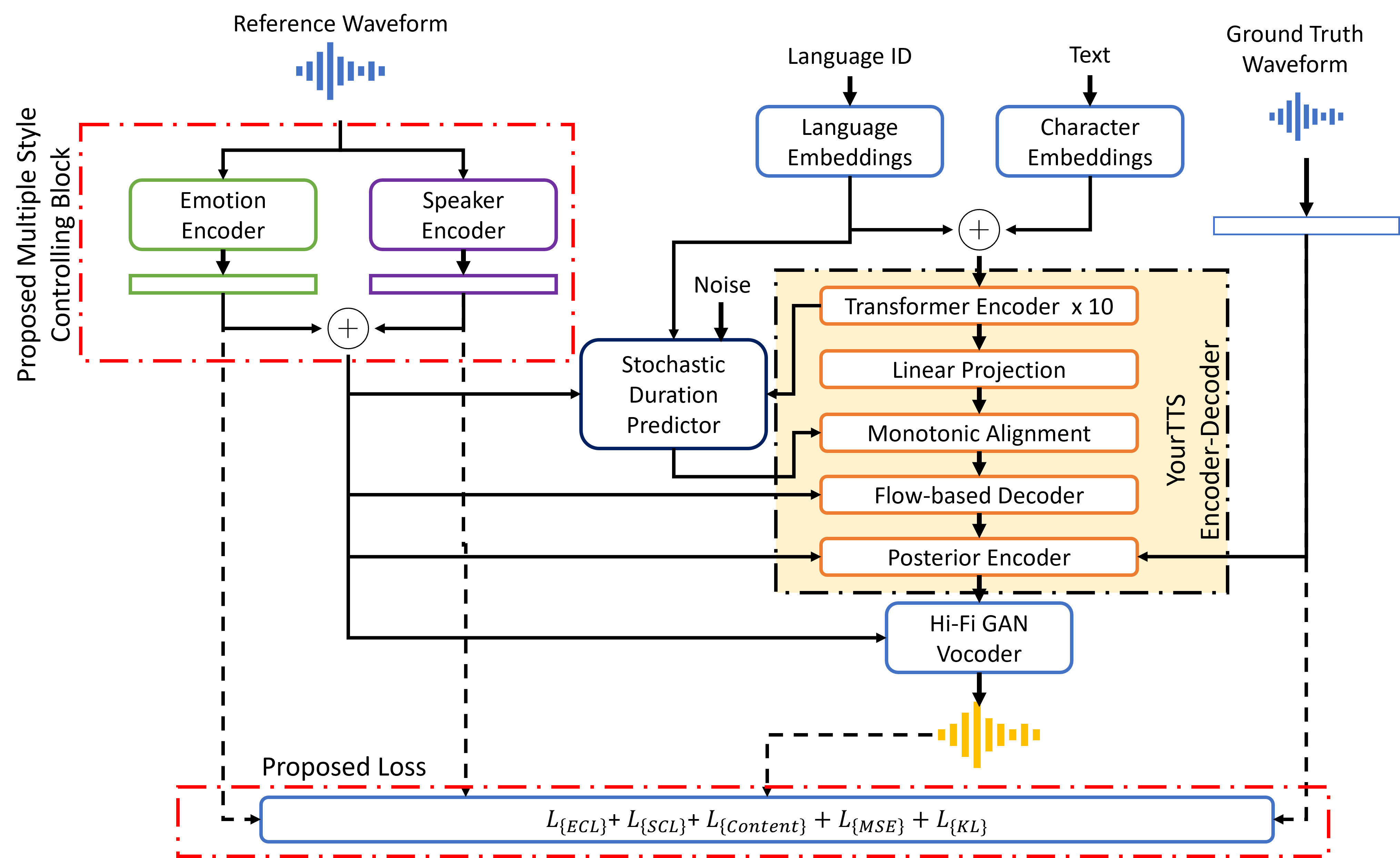}
    \vspace{-0.2cm}
    \caption{The bock diagram of proposed VECL-TTS model. Contributions are highlighted via red dotted box.}
    \vspace{-0.4cm}
    \label{fig:block_diag}
\end{figure*}
\subsection{Proposed VECL-TTS Model}
Our work builds upon the foundation of YourTTS \cite{casanova2022yourtts}. The proposed model uses raw text as input due to the unavailability of decent grapheme-to-phoneme converters for Indian languages. The model demonstrates its ability to synthesize good-quality speech directly from raw text and eliminates the need for the linguistically taxing job of creating suitable grapheme-to-phoneme converters. As shown in Figure \ref{fig:block_diag}, our model employs a transformer-based text encoder \cite{kim2020glowtts, vaswani2023attention} with \textit{10} transformer blocks and the number of hidden channels as 196. Moreover, we concatenate a 4-dimensional language embedding into each input character embedding to facilitate multilingual control. By doing so, we are conditioning the text encoder to use language information at every character, thus ensuring more stable outputs. For the decoder of the text module, we use a stack of 4 affine coupling layers \cite{dinh2017density}, where each layer is composed of 4 WaveNet residual blocks \cite{oord2016wavenet} following previous work \cite{casanova2022yourtts}.\\
\indent For generating the output speech waveform, we use HiFi-GAN V1 \cite{kong2020hifigan} as the neural vocoder. We have a Variational Auto-Encoder (VAE) \cite{kingma2022autoencoding} that uses a Posterior Encoder \cite{kim2021conditional}, which is composed of 16 WaveNet residual blocks \cite{oord2016wavenet} to convert linear spectrogram into a latent variable. This latent variable is fed into the vocoder and the flow-based decoder thus connecting the vocoder to the TTS model, making it end-to-end. The usage of learned latent variables helps the model to come up with its own representation of speech and eliminates the need for a mel-spectrogram. Having the model learn the intermediate representation of speech achieves better results as the representation learned might be more suitable for the task than traditional approaches.\\
\indent We use Stochastic Duration Predictor (SDP) \cite{kim2021conditional}, to generate speech with diverse rhythms from input text. The SDP is composed of Dilated Depth Separable Convolutions (DDSC) based Spline Flows \cite{durkan2019neural} and predicts the duration of each character, which is then used to generate the output in parallel, giving it faster than real-time inference. We propose multiple-style emotion control, which allows us to generate speech with different emotions even in a cross-lingual setup. We used a multi-lingual emotion encoder to get emotion embeddings for emotion controllability. Using these embeddings, we condition the YourTTS encoder-decoder to produce emotional speech, as explained in the following sections.
We also have a KL divergence loss (which we represented as, $L_{KL}$) because of the model's VAE nature  \cite{kim2021conditional,casanova2022yourtts}. This loss tries to maximize the evidence lower bound (ELBO) to get a desired multivariate normal distribution.
\subsection{Emotion Style Control}
Transferring speaker and emotion traits from one language to another is tricky. Prosody patterns and small nuances vary among languages due to differences in pronunciation.
In the proposed approach, we trained a multilingual emotion encoder to obtain the language-specific emotion representation, as shown in Figure \ref{fig:block_diag}. The motivation for using a pre-trained emotion encoder is to get more diverse emotion information representation from reference audio rather than using just emotion ID or one-hot encoding. 
The proposed multilingual emotion encoder is developed by training an emotion recognition model consisting of a parallel 2D Convolutional Neural Network (CNN) and a transformer encoder \cite{kosta2022emotion}.  This emotion recognition model is trained using multilingual data to learn language-independent representations. This model obtained an F1 score of 91.28\% using a multi-lingual test dataset. The embedding obtained from the emotion encoder represents emotion characteristics, which are used to condition the encoder, decoder, and duration predictor to generate emotional audio.

\par We also added Emotion Consistency Loss (ECL) to preserve targeted emotion in generated audio. To do so, we computed emotion embeddings from ground truth audio and generated audio, and maximized the cosine similarity between them. The ECL loss (as shown in Fig. \ref{fig:block_diag}) is defined in Eq. \ref{eq:ecl_loss} as follows,
\begin{equation}
    L_{ECL} = \frac{-\alpha_e}{n}\sum_{i}^{n}cos\_sim(\phi_e(gen_{i}), \phi_e(gt_{i})),
    \label{eq:ecl_loss}
\end{equation}
where $\phi_e(gen)$ and $\phi_e(gt)$ are functions to obtain emotion embeddings from generated and ground truth audio. The $\alpha_e$ is the tuning parameter used while adding ECL to the final loss to decide ECL's contribution to the final loss. The cosine similarity function is represented by $cos\_sim$.

\subsection{Speaker Identity Control}
In order to transfer speaker characteristics from one language to another, we conditioned the proposed TTS model using speaker embeddings obtained from the speaker encoder. The H/ASP \cite{heo2020clova} model is used as the speaker encoder as shown in Figure \ref{fig:block_diag}. We also used Speaker Consistency Loss (SCL) in the final loss. The SCL is obtained by maximizing cosine similarity between extracted embeddings from ground truth and generated audio, as shown in Eq. \ref{eq:scl_loss}.
\begin{equation}
    L_{SCL} = \frac{-\alpha_s}{n}\sum_{i}^{n}cos\_sim(\phi_s(gen_{i}), \phi_s(gt_{i})).
    \label{eq:scl_loss}
\end{equation}
\par In addition to ECL and SCL, we also added content loss \(L_{Content}\) to the final loss including mean squared error (MSE) loss (i.e., $L_{MSE}$) and $L_{KL}$. The content loss is the loss between wav2vec embeddings obtained from ground truth and generated audio. The final loss is represented by the Eq. \ref{eq:final_loss}.
\begin{equation}
    L_{final} = L_{ECL} + L_{SCL}+L_{Content}+L_{MSE}+L_{KL}.
    \label{eq:final_loss}
\end{equation}
\section{Evaluation}
\subsection{Experimental Database}
\label{sec:datasets}
In this work, we considered four languages, English, Hindi, Marathi, and Telugu, having one dataset for each language. We used the VCTK dataset for the English language, which comprises of 44 hours of speech from 109 different speakers \cite{veaux2016superseded}. For Indian language databases, we used the LIMMITS challenge \cite{zhang2023leanspeech,du2023multi} database. Each Hindi, Marathi, and Telugu data was recorded by two speakers. The total duration of the recordings in each language varies between 45-55 hours approximately. Additionally, we have used ~57.42 hours of internal in-house Emotional data from 11 speakers in the Hindi language. We considered six emotions: Neutral, Angry, Happy, Sad, Fear and Surprise. For all considered datasets, we performed pre-processing to remove long silences and loudness normalization to have similar loudness across the datasets\footnote{https://github.com/wiseman/py-webrtcvad}\footnote{https://github.com/slhck/ffmpeg-normalize}. Total data is divided into subsets of 80\% training, 10\% validation and 10\% for test. 
\subsection{Experimental Setup}
The model was trained on a single NVIDIA A100 80GB PCIe with a batch size of 128. For both the TTS module and HiFi-GAN vocoder, we used AdamW Optimizer \cite{loshchilov2019decoupled} with betas 0.8 and 0.99, weight decay 0.01 and an initial learning rate of 0.0002 decaying exponentially. The model was trained for 1M iterations and has around 88M parameters. 
\vspace{-0.3cm}
\subsection{SOTA Methods}
\begin{itemize}
     \item \textbf{YourTTS }\cite{casanova2022yourtts} is an effective multilingual TTS model. It uses a speaker encoder to generate speaker embeddings to condition the TTS model to achieve cross-lingual voice cloning.  
     \item \textbf{M3} \cite{liu4564500m} TTS is a multi-modal and multi-scale style TTS. Adversarial training and Mixed Density Networks (MDN) are used to map style vectors for voice cloning.
     \item \textbf{METTS} \cite{zhu2023metts} is a multi-lingual emotional TTS model which supports cross-lingual emotion and speaker transfer. It uses delightfulTTS as the backbone and introduces multi-scale emotion modeling.
     \item \textbf{CET} \cite{wu2021cross} TTS transfers emotions across speakers. Emotion tokens are trained to be closely related to corresponding emotions. The model transfers emotion from a reference mel-spectrogram. 
\end{itemize}
\vspace{-0.3cm}
\section{Results and Discussion}
This section presents the comparison and discussion between the proposed method and other SOTA methods. 
\vspace{-0.3cm}
\subsection{Subjective Evaluation}
In this work, performance evaluation is carried out to analyze the generated cross-lingual emotional speech quality in terms of naturalness, emotion, and speaker similarity. The Mean Opinion Score (MOS) \cite{ribeiro2011crowdmos} is used to evaluate the quality of synthesized speech. Total 20 listeners (aged between 23 to 35 with no known hearing impairments) participated in the subjective evaluation. MOS score is calculated by considering a total of 800 samples during each subjective test for naturalness, speaker similarity and emotional similarity. 
\begin{figure*}[ht]
    \centering
    \includegraphics[width=0.8\linewidth]{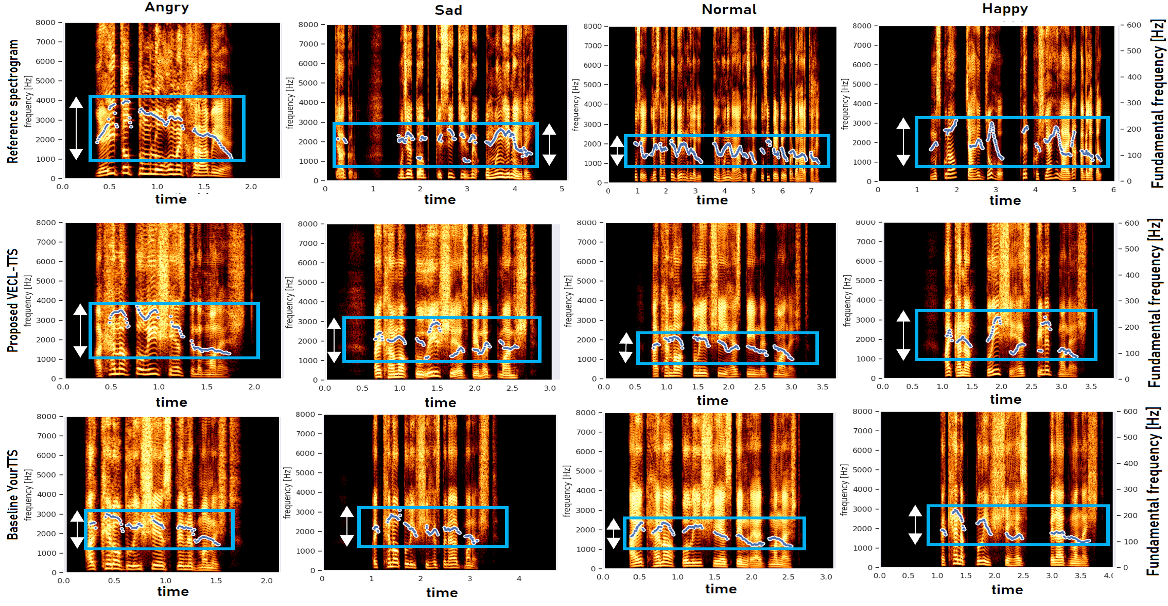}
    \vspace{-0.3cm}
    \caption{Visualization of Mel spectrogram and pitch variation for ground truth, proposed VECL-TTS and YourTTS generated speech.}
     \vspace{-0.3cm}
    \label{fig:pitch_var}
    \vspace{-0.3cm}
\end{figure*}
Table \ref{tab:sub_comp} shows the similarity MOS scores for the proposed model and SOTA approaches. The evaluation results show that the proposed model performs better than the baseline models for emotion similarity. Regarding naturalness, the proposed model performs better compared to all baselines except for METTS, where the MOS score is close to METTS. For speaker similarity score, the YourTTS model performs better. From Table \ref{tab:sub_comp}, we can say that the proposed model performs better than all the baseline models for transferring cross-lingual emotion characteristics without much degradation in speaker similarity and naturalness.
\begin{table}[ht]
\vspace{-0.2cm}
\centering
\caption{Analysis of different subjective and objective Evaluation scores along with 95\% confidence interval and obtained $p$-value $<$ 0.001.}
\vspace{-0.3cm}
\label{tab:sub_comp}
\setlength{\tabcolsep}{1pt}
\scalebox{0.75}{
\begin{tabular}{|c|cccc|cc|}
\hline
\multirow{2}{*}{\textbf{Model}}                               & \multicolumn{4}{c|}{\textbf{Subjective Evaluation}}                                                                                      & \multicolumn{2}{c|}{\textbf{Objective Evaluation}}                     \\ \cline{2-7} 
& \multicolumn{1}{c|}{\textbf{Quality}} & \multicolumn{1}{c|}{\textbf{Emo Sim}} & \multicolumn{1}{c|}{\textbf{Spk Sim}} & \textbf{Avg} & \multicolumn{1}{c|}{\textbf{Cos Emo Sim}} & \textbf{Cos Spk Sim} \\ \hline
YourTTS                                                       & \multicolumn{1}{c|}{3.06 $\pm$ 0.11}             & \multicolumn{1}{c|}{3.52 $\pm$ 0.11 }             & \multicolumn{1}{c|}{\textbf{3.95 $\pm$ 0.14}}    & 3.51             & \multicolumn{1}{c|}{0.70}                    & 0.85                    \\ 
METTS                                                         & \multicolumn{1}{c|}{\textbf{3.95 $\pm$ 0.12}}    & \multicolumn{1}{c|}{3.44 $\pm$ 0.16}             & \multicolumn{1}{c|}{3.82 $\pm$ 0.14}             & 3.73             & \multicolumn{1}{c|}{0.54}                       & 0.75                    \\ 
M3                                                            & \multicolumn{1}{c|}{2.41 $\pm$ 0.19}             & \multicolumn{1}{c|}{2.81 $\pm$ 0.18}             & \multicolumn{1}{c|}{3.24 $\pm$ 0.15}             & 2.82             & \multicolumn{1}{c|}{0.56}                       & 0.67                    \\ 
CET                                                           & \multicolumn{1}{c|}{2.89 $\pm$ 0.14}             & \multicolumn{1}{c|}{3.21 $\pm$ 0.12}             & \multicolumn{1}{c|}{3.33 $\pm$ 0.11}             & 3.14             & \multicolumn{1}{c|}{0.51}                       & 0.63                    \\ 
Ablation 1                                                    & \multicolumn{1}{c|}{3.2 $\pm$ 0.12}              & \multicolumn{1}{c|}{3.64 $\pm$ 0.11}             & \multicolumn{1}{c|}{3.64 $\pm$ 0.14}             & 3.49             & \multicolumn{1}{c|}{0.64}                    & 0.81                    \\ 
Ablation2                                                     & \multicolumn{1}{c|}{3.22 $\pm$ 0.11}             & \multicolumn{1}{c|}{3.42 $\pm$ 0.12}             & \multicolumn{1}{c|}{3.91 $\pm$ 0.11}             & 3.51             & \multicolumn{1}{c|}{0.60}                    & 0.83                    \\ 
\begin{tabular}[c]{@{}c@{}}VECL-TTS\\ (Proposed)\end{tabular} & \multicolumn{1}{c|}{3.85 $\pm$ 0.12}             & \multicolumn{1}{c|}{\textbf{3.82 $\pm$ 0.11}}    & \multicolumn{1}{c|}{3.81 $\pm$ 0.12}             & \textbf{3.82}    & \multicolumn{1}{c|}{\textbf{0.72}}           & \textbf{0.86}           \\ \hline 
\end{tabular}
}
\vspace{-0.17cm}
\end{table}
\par The CET \cite{wu2021cross} model is designed for transferring emotions across different languages, but it struggles to capture the various emotional expressions. This leads to synthetic speech with a strong accent and lower scores in naturalness and speaker similarity evaluations. In contrast, our proposed model uses emotion embedding to capture both language-specific and language-agnostic emotional expressions, avoiding accent-emotion entanglement. M3 \cite{shang2021incorporating}, on the other hand, performs poorly in all evaluation metrics. It assumes a strong correlation between style coding, speaker attributes, and content, resulting in an entanglement between the speaker's timbre and emotion. In contrast, our proposed model utilizes an emotion embedding network to effectively remove the speaker's timbre, resulting in a more natural emotional speech. Demo samples are available at demo page\footnote{\url{https://nirmesh-sony.github.io/VECL_TTS/}.}
\subsection{Objective Evaluation}
The objective evaluation is carried out to comprehensively evaluate the performance of the proposed VECL-TTS system. In objective evaluation, we computed cosine similarity score for speaker and Emotion similarities. In particular, cosine distance is computed between the speaker embeddings of the synthesized audio and ground truth audio. Similarly, the cosine distance between emotion embedding of synthesized audio and ground truth audio is computed for emotion similarity. We used a speaker and emotion encoder network for estimating style embeddings. A higher number of cosine similarities denotes stronger resemblance \cite{cooper2020zero}. 
Table \ref{tab:sub_comp} denotes the objective evaluation comparison between the proposed approach and baseline methods. The objective test results correlate well with subjective evaluations. Table \ref{tab:sub_comp} shows that the proposed VECL-TTS model achieves the highest score for cosine similarity compared to the SOTA approaches. The CET approach achieves a lower score for cosine speaker similarity, which indicates a significant challenge in transferring emotion characteristics in the cross-lingual scenario as it focuses on inter-lingual emotional TTS. Additionally, M3 fails to address accent-related challenges in multilingual emotional speech synthesis effectively.
\subsection{Ablation Study}
We perform ablation analyses, individually excluding the ECL and content loss. Ablation 1 denotes the YourTTS model, including content loss (i.e., proposed model w/o ECL) with the proposed multiple style controlling block. Similarly, Ablation 2 denotes the YourTTS model, including ECL loss (i.e., proposed model w/o content loss). Table \ref{tab:sub_comp} displays the corresponding ablation analysis subjective MOS score. Table \ref{tab:sub_comp} shows that when we add ECL loss to the baseline YourTTS, emotion similarity, and naturalness improve, as indicated in Ablation 2. Content loss in Ablation 1, which focuses on improving pronunciation, is evident from Table \ref{tab:sub_comp}, where it improves the naturalness of generated audio. 
It can be observed from the MOS scores of the proposed VECL-TTS and baseline YourTTS that while transferring emotion characteristics, the proposed VECL-TTS model maintains speaker similarity close to the baseline YourTTS.
\vspace{-0.3cm}
\subsection{Visual Analysis of VECL model}
Further, we visualize the changes in prosody patterns such as Mel spectrogram and pitch for different emotions between reference, generated, and YourTTS audio in a cross-lingual setting. Figure \ref{fig:pitch_var} shows that the generated audio using VECL-TTS for a particular emotion follows a similar trend in pitch variations compared to reference audio. For example, for angry and happy emotions, the range of pitch variation indicated by the blue box is higher for reference audio, and a similar pattern is found in the proposed VECL-TTS generated audio. In YourTTS, the pitch variations are much less when compared to reference audio. In contrast, VECL-TTS captures diverse pitch variations and is very similar in range compared to reference audio. We also notice that in YourTTS, the pitch variances in angry and happy are very similar to sad. This indicates that the proposed VECL-TTS model is more expressive when compared to the YourTTS model.
\vspace{-0.3cm}
\section{Summary and Conclusion}
In this paper, we proposed the VECL-TTS model to control the voice identity and emotion style simultaneously in a cross-lingual context. Here, we employed speaker and emotion embeddings from two multilingual pre-trained classifiers to obtain more refined style representations. Additionally, we introduced the ECL loss to preserve emotions. To address challenges related to pronunciation and degradation in generated speech resulting from cross-lingual style transfer, we proposed a content loss framework that utilizes wave2vec2-based self-supervised speech representations. The efficacy of our proposed VECL-TTS model is compared to the SOTA architectures within the framework of English and three Indian languages: Hindi, Telugu, and Marathi. We have observed that the proposed model attains better subjective and objective scores, particularly regarding emotional similarity compared to the pertinent baseline. In the future, we would like to further improvise the output with respect to naturalness, speaker, and emotion similarity simultaneously.
\bibliographystyle{IEEEtran}
\balance
\bibliography{custom}

\end{document}